\documentstyle[eqsecnum,aps,epsf,epsfig]{revtex}

\newcommand{\be}{\begin{eqnarray}} \newcommand{\ee}{\end{eqnarray}}

\newcommand{\lton}{\mathrel{\lower.9ex \hbox{$\stackrel{\displaystyle<}{\sim}$}}}

\newcommand{\nn}{\nonumber} 
\newcommand\del{\partial}

\newcommand{\mat}{\left ( \begin{array}{cc}}
\newcommand{\emat}{\end{array} \right )} \newcommand{\matt}{\left (
\begin{array}{ccc}} \newcommand{\ematt}{\end{array} \right )}
\newcommand{\matf}{\left ( \begin{array}{cccc}}
\newcommand{\ematf}{\end{array} \right )} \newcommand{\vect}{\left (
\begin{array}{c}} \newcommand{\evect}{\end{array} \right )}
\newcommand{\Tr} {\rm Tr} 
\def\beqn{\begin{eqnarray}} \def\eeqn{\end{eqnarray}} 
 \def\om{\omega}  \def\d{\partial}
\def\Tr{ {\rm Tr} }

\begin{document}

\vspace{1cm}

\title{
\begin{flushright} 
{\small SUNY-NTG-02/32}
\end{flushright} 
 Fluctuation Induced Critical Behavior at Nonzero Temperature
\\  and Chemical Potential}  
\author{K.\ Splittorff$^{\, (a)}$, J.\ T.\ Lenaghan$^{(b)}$, and 
J.\ Wirstam$^{(c)}$\footnote{Former address: Department of 
Physics, Brookhaven National Laboratory, Upton, {\sl New York 11973}}}
\address{
$^{(a)}$ Department of Physics and Astronomy, SUNY, Stony
Brook, {\sl New York 11794}\\ $^{(b)}$ Department
of Physics, University of Virginia, 382 McCormick Rd.,
Charlottesville, {\sl VA 22904-4714} \\ $^{(c)}$ Swedish Defense
Research Agency (FOI), S-172 90 Stockholm, Sweden}

\maketitle

\begin{abstract}
We discuss phase transitions in relativistic systems as a function of
both chemical potential and temperature.  The presence of a chemical
potential explicitly breaks Lorentz invariance and may additionally
break other internal symmetries.  This introduces new subtleties in
the determination of the critical properties.  We discuss separately
three characteristic effects of a nonzero chemical potential.
Firstly, we consider only the explicit breaking of Lorentz invariance
using a scalar field theory with a global $U(1)$ symmetry. Secondly,
we study the explicit breaking of an internal symmetry in addition to
Lorentz invariance using two--color QCD at nonzero baryonic chemical
potential.  Finally, we consider the spontaneous breaking of a
symmetry using three-color QCD at nonzero baryonic and isospin
chemical potential.  For each case, we derive the appropriate 
three-dimensional effective theory at criticality and study the effect of
the chemical potential on the fixed point structure of the
$\beta$-functions. We find that the order of the phase transition is
not affected by the explicit breaking of Lorentz invariance but is
sensitive to the breaking of additional symmetries by the chemical
potential.
\end{abstract}


\section{Introduction}
\label{sec:intro}

The concept of universality is a powerful tool in the description of
second--order phase transitions.  In many cases, a mean field
treatment of the Landau theory for the order parameter provides a
sufficient description of the universal properties of the phase
transition.  However, in some cases, a mean--field analysis is not
reliable since higher-order corrections to the Landau theory may
result in $\beta$-functions which do not have stable fixed points.
The form of the higher-order corrections depends upon the detailed
symmetries of the problem.  In particular, higher-order corrections to
the $\beta$-function may drive a phase transition which is
second--order at the level of mean--field theory to be first order on
account of quantum fluctuations.  Such phase transitions are called
fluctuation induced phase transitions.  The mechanism by which this
occurs at the level of the $\beta$-function is simple to understand.
A second--order phase transition requires the existence of nontrivial
stable fixed points for the $\beta$-function since such a transition
is characterized by a diverging correlation length.  Quantum
corrections may drive this fixed points towards instability or even
make it vanish \cite{BKM,BD,amit,cw,yamagishi}.

In this article, we discuss the influence of a nonzero chemical
potential on the fixed point structure of $\beta$-functions for
systems which are relativistically invariant at zero temperature and
zero chemical potential. Both the temperature and the chemical
potential explicitly break the O(4) relativistic invariance to
O(3).\footnote{Throughout this article, we will work in Euclidean
space.}  The presence of a chemical potential may also break
additional symmetries both explicitly and spontaneously through, for
example, Bose condensation or Cooper pairing. Hence, there are (at
least) three characteristic effects of the chemical potential:
\begin{itemize}
\item{the explicit breaking of Lorentz invariance,} 
\item{the explicit breaking of internal symmetries in addition to Lorentz invariance,}
\item{the spontaneous breaking of symmetries.}
\end{itemize}

We consider three different theories to illustrate the physics issues
associated with each of the three above effects.  First we consider a
complex scalar field theory with a global $U(1)$ symmetry. We
introduce a chemical potential for this $U(1)$ charge. This allows
for additional kinetic terms in the Landau effective theory and these
kinetic terms break Lorenz invariance explicitly.  In the $U(1)$
model, no additional symmetries are explicitly broken and hence it is
ideal to illustrate the first point above. In particular, we use this
example to illustrate dimensional reduction from four to three
dimensions at nonzero temperature and chemical potential.

To exemplify the second and third points we consider the chiral
symmetry restoring phase transition in massless QCD with two and three
colors, respectively.  The chemical potential may alter the pattern of
chiral symmetry breaking which in turn changes the spectrum of
Goldstone excitations.  We examine how this change in the chiral
symmetry breaking pattern affects the order of the temperature induced
chiral phase transition as a function of chemical potential.  The
chiral order parameters discussed here are all color singlets. The
possibility of superconducting phases has been studied using
renormalization group techniques in \cite{BCSreno}. In that case gauge
fields can play an important role and are incorporated into a
Landau-Ginzburg theory.

To be more specific, consider two--color QCD. In this case, a baryon
chemical potential breaks the $U(2N_{\! f})$ classical chiral symmetry
explicitly to $U(N_{\! f})\times U(N_{\!f})$.  In the hadronic low
energy spectrum, this change in the chiral symmetry breaking pattern
manifests itself in that the Goldstone bosons have different baryon
charges. At low energy, the Goldstone modes are weakly coupled and hence
correctly charged ones will form bose condensates when the baryon
chemical potential is sufficiently high and the temperature
sufficiently low. The formation of this Bose condensate breaks 
$U(N_{\! f})\times U(N_{\!f})$ spontaneously to 
$Sp(N_{\! f})\times Sp(N_{\!f})\times U(1)$.  As the temperature 
increases the Bose condensate melts and chiral symmetry is restored. 
We study the order of this phase transition. See figure
\ref{fig:2colmq=0} for an illustration.

The case of three--color QCD is somewhat special in that the baryonic
chemical potential couples only indirectly to the effective theory
since the chiral order parameter is not charged under baryon number.
As such, the presence of a baryonic chemical potential does not
violate the usual chiral symmetry breaking pattern $SU(N_{\! f})\times
SU(N_{\!f}) \rightarrow SU(N_{\!f})$.  This in turn implies that the
$\beta$-functions are not affected by the presence of a baryonic
chemical potential. This point was previously raised in \cite{HS}, and
here we discuss the region in the $(\mu,T)$-plane where such a
simplification applies.

Finally we consider a chemical potential for the third component of
isospin in three-color QCD with two light quark flavors. Here the
situation is almost identical to the one described for two--color QCD
at nonzero baryon chemical potential. An isospin chemical potential
explicitly breaks chiral symmetry. The pions form an isospin triplet
and with increasing isospin chemical potential a pion condensate
forms \cite{SS}. This condensate breaks chiral symmetry at low temperature but
as the temperature is raised chiral symmetry is restored.  We study
the order of this phase transition using the $\epsilon$-expansion.

The approach followed in this article begins with the
four--dimensional Landau theory for the relevant order parameter
$\Phi$. 
This effective theory is the most general renormalizable
Lagrangian consistent with the relevant symmetries.  Because of the
explicit breaking of Lorentz invariance the kinetic term may take a
nonstandard form ($B$ is the charge matrix defined below): 
\be
\Tr\left[\d_0\Phi^\dagger\d_0\Phi\right] 
+v^2(\mu,T)\Tr\left[\d_i\Phi^\dagger\d_i\Phi\right]+\mu
q_1(\mu,T)\Tr\left[B\Phi^\dagger\d_0\Phi\right]+ \mu q_2(\mu,T)\Tr\left[B\Phi\d_0\Phi^\dagger\right] 
\ .  
\ee 
The functions $q_1$ and $q_2$ are constrained by the symmetries of the
underlying microscopic Lagrangian and determine the conserved current
for the charge to which the chemical potential is associated.  This is
analogous to how the chemical potential enters in chiral perturbation
theory \cite{KST,KSTVZ,SS,SSS,KT,LSS,STV1,STV2}, the difference being
mainly that the Goldstone field has fewer components than the
generalized order parameter field of the Landau theory.  Before we
start our analysis let us emphasize that the presence of fixed points
only implies that the phase transition can be of second order, since
higher dimensional operators may render the phase transition first
order. Additionally, the possible existence of a nonperturbative fixed
point may drive a phase transition to second order.
Let us also point out that in addition to their intrinsic interest, 
our results should be useful
for the interpretation of lattice gauge theory results.  In general,
lattice gauge theory simulations at nonzero chemical potential suffer
from the notorious sign problem.  Two--color QCD at nonzero baryonic
and isospin chemical potential and three-color QCD at nonzero isospin
chemical potential have positive measures and so are special cases of
dense gauge theories which can be simulated by standard Monte--Carlo
methods.  Hence it is possible to conduct first principle numerical
computations at nonzero isospin chemical potential using the standard
methods \cite{AKW}.
Such simulations have exposed \cite{Kogut:2002tm,KS2} a rich phase
diagram in the $(\mu_I,T)$-plane. For $T=0$ the pion condensate sets
in when $\mu_I$ exceeds the pion mass. This transition into the pion
phase is second order. For temperatures on the order of the pion mass
and higher the phase transition changes from second order to first
order. This scenario has been explained within the context of chiral
perturbation theory \cite{STV2}.  Making a direct comparison to the
predictions in the present paper is delicate since the lattice
simulations necessarily work at a nonzero quark mass. 

The paper is organized as follows. In the next section, we discuss the
case where the introduction of a chemical potential only leads to a
breaking of the Lorentz invariance (or, in Euclidean space, of the
O(4) rotational symmetry). In sec.\ III we extend the analysis to 
$N_{\! c}=2$ QCD, where, as mentioned above, additional 
global flavor symmetries are broken. Three-color QCD is discussed
in sec.\ IV and sec.\ V, and we end with our conclusions in sec.\ VI.

\section{When $\mu$ breaks only Lorentz invariance}
\label{sec:dimreduct}

In this section, we consider a theory in which the chemical potential
does not break any internal symmetries.  The simplest possible example
is the given by a complex scalar field theory with a $U(1)$ symmetry
to which a chemical potential is coupled. This example extends the one
given in \cite{ginsparg} to nonzero chemical potential. The
Lagrangian is  
\be\label{LagrangianU(1)} 
L_{4d}  & = &
\del_0\Phi^*\del_0\Phi+v^2\del_i\Phi^*\del_i\Phi+\mu \left[\Phi\del_0
\Phi^*-(\del_0
\Phi)\Phi^*\right]+(m^2-\mu^2)\Phi^*\Phi+\lambda(\Phi^*\Phi)^2 \ . 
\ee 
The Lagrangian reveals the standard coupling of $\mu$ to the
zeroth component of the conserved current. It is the lowest order
coupling of the chemical potential to the order parameter field
consistent with the $U(1)$ invariance.

The analysis of the critical behavior proceeds in
four steps: 
\begin{itemize}
\item{1.) Fourier decompose the fields and integrate over
$x_0\in [0,1/T]$.}  
\item{2.) Determine the propagators in the resulting three--dimensional theory.}
\item{3.) Integrate out the massive Matsubara modes to get the
effective three--dimensional theory.}
\item{4.) Study the stability of the fixed points of the
$\beta$-functions in this three--dimensional effective theory.}
\end{itemize}

Writing out the Lagrangian in Eq.\ (\ref{LagrangianU(1)}) in terms of
the real components of the order parameter, $\Phi\equiv a+ib$, we find 
\be
L_{4d} & = & \del_0 a\del_0 a+v^2\del_i a\del_i a  +\del_0 b\del_0
b+v^2\del_i b\del_i b - 2i\mu \,
\left(a\del_0b-b\del_0 a\right)
+\left(m^2-\mu^2\right)\left(a^2+b^2\right) +\lambda \,
\left(a^2+b^2\right)^2 \ .  
\ee 
Since the temporal direction is
compact, we may Fourier decompose the fields as
\be 
a(x_0,\vec{x}) = T \sum_{n=-\infty}^\infty e^{i\omega_n x_0} a_n(\vec{x}) \ \ \ \ {\rm and}
 \ \ \ \ b(x_0,\vec{x}) = T \sum_{n=-\infty}^\infty e^{i\omega_n x_0} b_n(\vec{x}) \,\,.
\ee 
Dimensionally reducing this theory amounts to inserting these
expressions into the action and integrating $x_0$ from 0
to $1/T$. Using the fact that $a_{-n}=a_n^*$ and $b_{-n}=b_n^*$, we get 
\be
L_{3d} & = & T \sum_{n=-\infty}^\infty \left[v^2\del_i a_n\del_i a_n^*
+v^2\del_i b_n\del_i 
b_n^*+\left\{\omega_n^2+m^2-\mu^2\right\}(a_na_n^*+b_nb_n^*)
-2\mu\omega_n(a_nb_n^*-b_na_n^*)\right]+\lambda-{\rm terms} \ .   
\ee 
Next, we write the three-dimensional Lagrangian
in terms of the real fields $a_n\equiv c_n+id_n$ and $b_n\equiv
e_n+if_n$: 
\be 
L_{3d} & = & T \sum_{n=-\infty}^\infty \big[v^2((\del_i c_n)^2+(\del_i d_n)^2+(\del_i e_n)^2
+(\del_if_n)^2)+\left\{\omega_n^2+m^2-\mu^2\right\}(c_n^2+d_n^2+e_n^2+f_n^2) \\
&& - 4 i \mu \omega_n (d_ne_n-c_nf_n)\big]+\lambda-{\rm terms} \
. \nonumber 
\ee 
The propagator matrix in the $(c_n,f_n)$-sector and
equivalently in the $(e_n,d_n)$-sector is 
\be\label{propU(1)}
\left( \matrix{ 
v^2 p^2+\omega_n^2+m^2-\mu^2 & -2i\mu\omega_n \cr
-2i\mu\omega_n & v^2 p^2+\omega_n^2+m^2-\mu^2} \right) \ .  
\ee

Given the propagator, we can derive the effective three--dimensional 
theory by integrating out the massive Matsubara modes which in the
present case are all the nonzero modes. In doing so we
extend Ginsparg's analysis \cite{ginsparg} to $\mu\neq0$. We first
choose the vacuum expectation value of the order
parameter to be in the direction of $a_0$, i.e. $b$ is the Goldstone
boson. The leading order contribution in $\lambda$ from the nonzero
Matsubara modes in the effective three--dimensional theory of the
massless zeroth Matsubara mode is then the one-loop correction to
the mass of the $a_0$. Schematically in going from four dimensions to
three dimensions we have
\be
 (m^2-\mu^2)a_0^2 \to (m^2-\mu^2)a_0^2+ \lambda T
\left(3\sum_{n\neq0}a_na_{-n}+\sum_{n\neq0}b_nb_{-n}\right)a_0^2 \ .
\ee
Using the propagator in Eq.\ (\ref{propU(1)}), each of the summations
over $n\neq0$ leads to a correction term 
\be M^2(T,m^2,\mu^2)
& = & \frac{1}{2}\lambda T\sum_{n\neq 0}\int \frac{{\rm d}^{d-1}
p}{(2\pi)^{d-1}}\left(\frac{1}{v^2 p^2+m^2-(\mu+i\om_n)^2}+
\frac{1}{v^2  p^2+m^2-(\mu-i\om_n)^2} \right) \ .
\ee 
Inserting 
\be 
\frac{1}{v^2 p^2+m^2-(\mu\pm i\om_n)^2} = \int_0^\infty {\rm d}t \
e^{-(v^2 p^2+m^2-(\mu\pm i\om_n)^2)t} 
\ee 
and integrating first over $p$ then over $t$ we get
\be 
M^2(T,m^2,\mu^2) & = & \frac{\lambda
T}{2}\frac{\pi^{(d-1)/2}}{(2\pi v)^{d-1}}\Gamma\left(\frac{3}{2}-\frac{d}{2}
\right)
\sum_{n\neq0}\left[(m^2-(\mu+i\om_n)^2)^{(d-3)/2}+(m^2-(\mu-i\om_n)^2)^{(d-3)/2}\right]
\ .  
\ee 

Assuming that $T\gg |m|$, we can expand the argument of the sum in
powers of $m^2/\om_n^2$. In the limit $d\to4$ and ignoring the
regularization of the next to leading order terms, we find
\be 
M^2(T,m^2,\mu^2) & = & -\frac{\lambda T^2}{4\pi v^3} \sum_{n=1}^\infty
2\pi n \left[2+\frac{1}{1+(\mu/\om_n)^2}\frac{m^2}{\om_n^2}
-\frac{1-3(\mu/\om_n)^2}{4(1+(\mu/\om_n)^2)^3}\frac{m^4}{\om_n^4}
+{\cal O}\left(\frac{m^6}{\om_n^6}\right)\right] \ .  
\ee 
The leading order term, $M^2(T,m^2,\mu^2) =  \lambda T^2/(12v^3)(1+{\cal
O}(m^2/(2\pi T)^2))$, is independent of both $m$ and $\mu$. Moreover,
this holds independently of the ratio\footnote{This is not the case if we
consider an imaginary chemical potential. In that case, the nonzero
Matsubara modes become massless when $\mu=\om_n$. Consequently, they
must be included in the effective theory for the phase transition.} of
$\mu$ and $T$.
In other words, at leading order in $m^2/(2\pi T)^2$, the correction takes
the same form as at $\mu=0$ and we find
\be\label{LU(1)zero-mode} 
L_{3d-{\rm eff}} =
\del_i a_0\del_i a_0 +\del_i b_0\del_i b_0+\left(m^2-\mu^2+\frac{\lambda
T^2}{3 v^3}\right)a_0^2+\lambda \left(a_0^2+b_0^2\right)^2 \ .  
\ee 
Hence, if the $U(1)$ symmetry is broken at zero temperature, we
reproduce the standard result for the symmetry restoration 
temperature\footnote{See, for example, Refs.\ \cite{HW,kapustaPRD,BBD} where the
relation to the effective potential is also explained in detail. The 
renormalization of the speed of light and the emergence of new
critical behavior associated with dynamics near the critical point is
discussed in Ref.\ \cite{BdeV}.}. 
Of course, for the theory to be self--consistent at $T_c$, the condition $T_c\gg |m|$ must be
fulfilled. From Eq.\ (\ref{LU(1)zero-mode}), we see that 
$T_c^2 = 3v^3(\mu^2-m^2)/\lambda$. If $m^2<0$ and $\lambda \ll 1$, the consistency
requirement is always fulfilled. If the coupling constant increases
to $\lambda\simeq 1$ while $m^2<0$, we still have $T_c\gg |m|$ if
$\mu^2\gg -m^2$. In this case, however, the one--loop approximation is no longer
trustworthy for such large values of $\lambda$, and the results not reliable. 
On the other hand, when $m^2>0$ the situation is
different. Since $T^2_c>0$, a necessary condition is $\mu \geq m$, and
unless $\lambda$ is {\em extremely} small, we must have
$\mu\gg m$ for consistency.

The point to be taken from this analysis is that the presence of the
chemical potential in the effective three--dimensional theory is felt
only through the direct term, $-\mu^2 \Phi^*\Phi$.  At criticality the
quadratic mass-like terms vanish, and the appropriate effective theory
in three dimensions at criticality contains only spatial derivative
terms and quartic couplings.  Specifically, in the $U(1)$--model
considered above, the effective Lagrangian is reduced to,
\be
L_{3d-{\rm eff}}(T_c) = \del_i a_0\del_i a_0 +\del_i b_0\del_i
b_0+\lambda(a_0^2+b_0^2)^2 \ .  
\ee 
The $\beta$-function for this theory is well studied \cite{GZ-J}. 
It has stable fixed points and consequently the phase transition is
second order. Again we want to emphasize that this holds true
independently of the value of $\mu$ as long as $T_c\gg |m|$. 

This ends our discussion of the $U(1)$ model. In conclusion: There 
is no effect of the explicit breaking of Lorentz invariance on the 
order of the phase transition as long as $T_c\gg |m|$.


\section{When $\mu$ breaks internal symmetries}
\label{sec:Nf=Nc=2}

We now consider the case in which the chemical potential explicitly
breaks symmetries in addition to Lorentz invariance. A good example is
provided by QCD with two colors and two flavors at nonzero baryonic
chemical potential.  For simplicity, we first ignore the effects of
the $U_A(1)$ axial anomaly.  At the end of this section, we state
the role of the axial anomaly on the order of the phase transition.  
A complete treatment of the structure
of the $\beta$-functions is given in Ref.\ \cite{WLS}. Here we explain 
in detail how the dimensional reduction assumed in \cite{WLS} comes
about.

\begin{figure}
\epsfxsize=4in
\centerline{\epsfbox{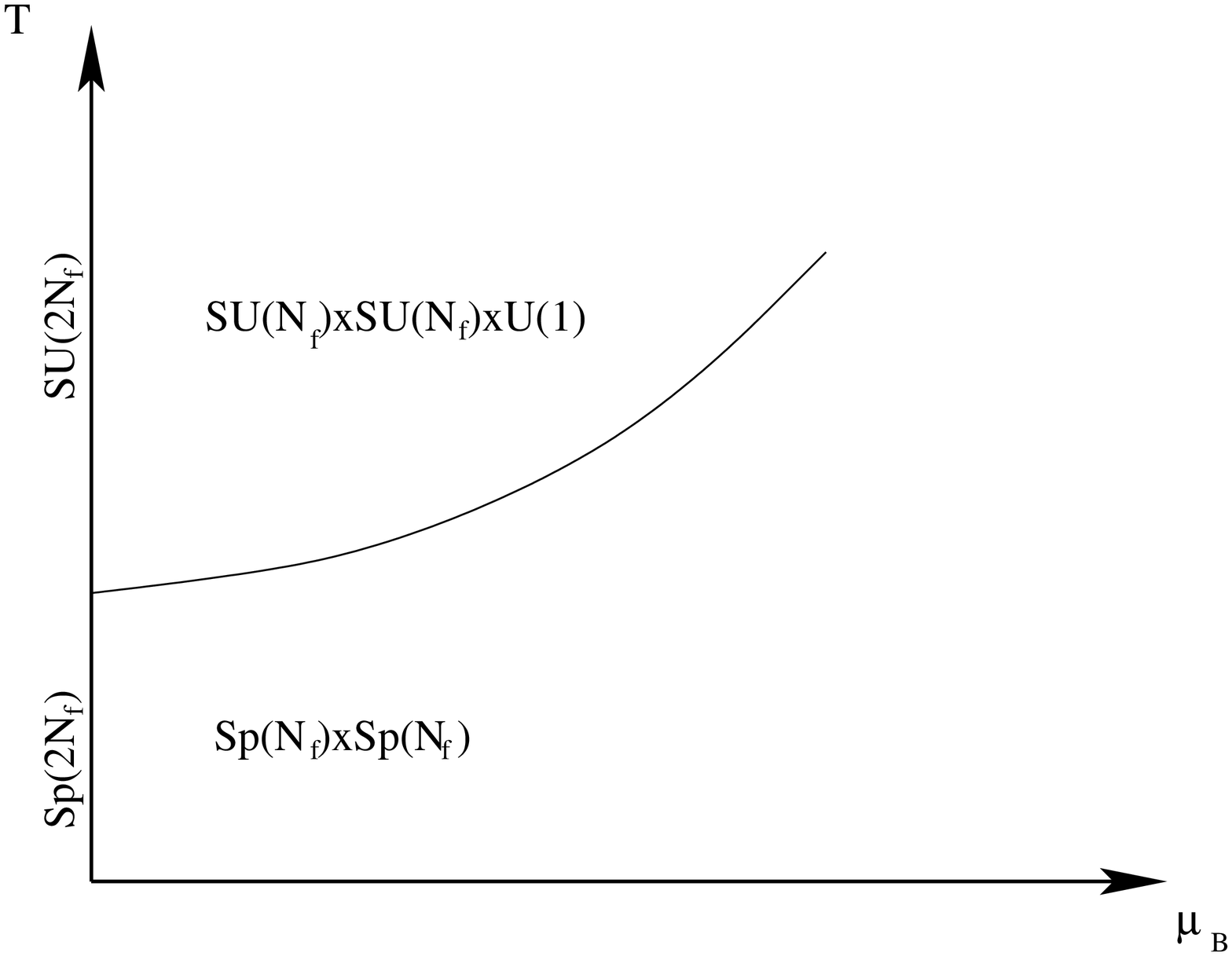}}
\caption{Symmetries in the temperature versus baryon
chemical potential phase diagram for massless two color QCD. The axial
anomaly is assumed to be present at all temperatures and chemical
potentials considered. \label{fig:2colmq=0}}
\end{figure}

QCD with two colors and two massless flavors enjoys a $U(2N_{\! f})$
classical invariance. This invariance is explicitly broken to 
$U(N_{\!f})\times U(N_{\! f})$ when $\mu\neq0$. This remaining symmetry is
spontaneously broken down to $Sp(N_{\! f})\times Sp(N_{\! f})\times U(1)$ 
for any nonzero value of $\mu$ by the formation of a diquark
condensate (see eg. \cite{KST,KSTVZ}). For illustration see figure
\ref{fig:2colmq=0}.  This diquark condensate signals  
the onset of a superfluid phase. We assume that the superfluid 
order parameter field can be represented by a
complex anti-symmetric matrix,
\be 
\Phi\equiv \left( \matrix{
X_1 & X_2 \cr -X_2^T & X_3} \right) \label{definematrix}
\ee 
where $X_1^T=-X_1$ and $X_3^T=-X_3$. In this basis, the baryonic
charge matrix is 
\be 
B\equiv \left( \matrix{1 & 0 \cr 0 & -1} \right) \ .
\ee 

The four--dimensional Landau theory for QCD with two
colors which is appropriate for the renormalization group analysis of the
temperature induced phase transition must be invariant under the same
$U(2N_{\! f})$ symmetry as the microscopic Lagrangian, and so is given
by (cf. \cite{KST,KSTVZ})
\be 
L_0 & = & 
\frac{1}{2}\Tr[(\del_\nu\Phi^\dagger ) (\del_\nu\Phi
)-2\mu\{\Phi,B\}\d_0\Phi^\dagger-2\mu^2(\Phi\Phi^\dagger+\Phi
B\Phi^\dagger B)] + V \ ,  
\ee 
where the potential is
\be \label{pot} V= \frac{m^2}{2} {\rm
Tr} \left [\Phi^{\dagger} \Phi \right ] +\lambda_1 \bigl ( \Tr \left [
\Phi^{\dagger} \Phi \right ] \bigr )^2 + \lambda_2 {\rm Tr} \bigl [
\left (\Phi^{\dagger} \Phi \right )^2 \bigr ]\ .  
\ee  
In the four--dimensional theory, the quadratic part of the Lagrangian
becomes, by using the representation in Eq.\ (\ref{definematrix}),  
\be 
L_{{\rm quad}} & = & \frac{1}{2}\Tr[(\del_{\nu}\Phi^{\dagger} )
(\del_{\nu}\Phi )]+\frac{m^2}{2} {\rm Tr} \left [\Phi^{\dagger} \Phi
\right ]-\mu^2{\rm Tr} \left [\Phi\Phi^\dagger+\Phi B\Phi^\dagger
B\right ]-\mu{\rm Tr} \left [\{\Phi,B\}\d_0\Phi^\dagger\right ] \\ 
& = & \frac{1}{2}\Tr[\del_\nu
X_1^\dagger\del_\nu X_1]+\Tr[\del_\nu X_2^\dagger\del_\nu
X_2]+\frac{1}{2}\Tr[\del_\nu X_3^\dagger\del_\nu X_3] \nn \\ 
& + & m^2 \Tr[X_2^\dagger X_2]+\frac{1}{2}(m^2-4\mu^2)\Tr[X_1^\dagger
X_1+X_3^\dagger X_3]-2\mu\Tr[X_1 \d_0 X_1^\dagger-X_3 \d_0
X_3^\dagger] \ .  \nn
\ee 
Note the explicit breaking of $U(2N_f)$ at $\mu\neq0$. 

We now proceed with the dimensional reduction of the theory as outlined 
in the previous section. For clarity, we will consider the specific
case of $N_f = 2$. From the equation
above, it is clear that the $X_2$ modes do not couple directly to the
chemical potential. The minimum of the potential for the diquark order
parameter 
\be
\Phi_0=\phi_0\left(\begin{array}{cccc}0&-i&0&0 \\ i&0&0&0 \\ 0&0&0&-i
\\ 0&0&i&0 \end{array}\right)
\ee
changes with $\mu$, however. This implies that the $X_2$ modes 
remain massive at the phase transition if $\mu\neq0$. Hence, all 
Matsubara modes from $X_2$ must be integrated out. The $X_1$ and $X_3$ modes do couple
to the chemical potential and we now derive their dispersion relations
in the three--dimensional theory. After setting 
\be 
X_1 \equiv
\frac{1}{\sqrt{2}}\left( 
\matrix{0 & a_1+i b_1 \cr -a_1- i
 b_1 & 0} \right) 
\ee 
and likewise for $X_3$, Fourier decomposing the real fields and
integrating over $x_0\in[1,1/T]$, we arrive at the three--dimensional theory.
The dispersion relations in the $X_1$ (or $X_3$) sector
are given by   
\be 
\left( \matrix{ 
p^2+\omega_n^2+m^2-4\mu^2 & -4i\mu\omega_n \cr
-4i\mu\omega_n & p^2+\omega_n^2+m^2-4\mu^2} \right) \ .  
\ee 
These are precisely the same dispersion relations as we found in the $U(1)$ 
model, eq. (\ref{propU(1)}), except that the charge of the modes is two 
and not one. When $m^2= (2\mu)^2$, there are four massless modes in
the three--dimensional theory, namely the four zeroth Matsubara modes of $X_1$
and $X_3$. In contrast, all modes in the $U(1)$ model are charged and all the zeroth
Matsubara modes are massless at criticality. In the
present case, only the $X_1$ and $X_3$ modes have nonzero baryonic charge
and these are exactly the fields with zeroth Matsubara modes that are massless
at criticality. This justifies the dimensional reduction 
applied in \cite{WLS}. The effective theory at the phase transition is given
in terms of these zeroth Matsubara modes and reads   
\be 
L_{3d-{\rm eff}}(T_c) & = & \frac{1}{2}\left[\d_j
a^{(1)}_0\d_j a^{(1)}_0 +\d_j a^{(3)}_0\d_j a^{(3)}_0 + \d_j
b^{(1)}_0\d_j b^{(1)}_0+ \d_j b^{(3)}_0\d_j b^{(3)}_0\right] \\ &&
+\lambda_1
\left[\left(a_0^{(1)}\right)^2+\left(b_0^{(1)}\right)^2+\left(a_0^{(3)}\right)^2+\left(b_0^{(3)}\right)^2\right]^2 \nn \\ && 
+\frac{\lambda_2}{2}\left[\left(\left(a_0^{(1)}\right)^2+\left(b_0^{(1)}\right)^2\right)^2+\left(\left(a_0^{(3)}\right)^2+\left(b_0^{(3)}\right)^2\right)^2\right]
\nn \ .  
\ee 
This Lagrangian has an $O(2)\times O(2)$ symmetry.
From the results of Ref.\ \cite{PS}, we know that the $\beta$-functions have
a marginally stable fixed point.  Hence, the order of the phase
transition is not determined at one-loop level.

If one takes the axial anomaly into account, the symmetry is reduced
to $O(2)$ and the $\beta$-function develops a fixed point
\cite{WLS}. Consequently, the phase transition is of second order in
the presence of the axial anomaly. At zero chemical potential and with
the anomaly present, the phase transition remains second order,
however, now with O(6) critical exponents \cite{wirstam}.

From this we conclude that the fixed point structure of the 
$\beta$-functions is affected by the
chemical potential only through the explicit breaking of the flavor
symmetries.  This result holds as long as $T_c\gg |m|$, and the
consistency requirement is fulfilled for all values of $\mu$ if
$m^2<0$. On the other hand, when $m^2>0$, self-consistency requires
that $\mu\gg m$.  The case $m^2<0$ mimics the phase structure for
massless quarks while the case with $m^2>0$ resembles two--color QCD
when the quarks have a common, nonzero mass. Some caution must be
exercised when considering infinitesimally small values of $\mu$,
since in that case the masses of $X_2$-modes in $\Phi$ are not well
separated from those of $X_1$ and $X_3$. The results given above only
apply when the mass scales are well separated. This ends our
discussion of two color QCD. We shall return to discuss the explicit
breaking of internal symmetries in addition to Lorentz invariance in
section \ref{sec:Nc=3muI}.


\section{Three--color QCD at nonzero baryonic chemical potential}
\label{sec:Nc=3}

We now consider the chiral phase transition in QCD with
three colors and massless quarks at nonzero baryonic chemical
potential. 
The baryon chemical potential does not break the
$SU(N_f)\times SU(N_f)$ chiral symmetry and if $\mu_B< M_{\rm Nucleon}/3$
no charged condensate is induced. Hence the order parameter remains
unaltered, i.e. the symmetry breaking pattern is unaltered and so is the
number of massless modes at the phase transition.
For larger chemical potentials, the behavior of the order parameter is
more complicated and we shall not discuss this case here. 
Following \cite{robfrank}, we assume that this chiral symmetry breaking 
order parameter field can be parametrized by a complex
$N_f\times N_f$ $\Phi$ for the values of $\mu_B$ under
consideration. Because the baryonic chemical potential 
explicitly breaks Lorentz invariance, the allowed form of the kinetic
term is  
\be
\Tr[\d_0\Phi^\dagger\d_0\Phi]+v^2(\mu,T)\Tr[\d_i\Phi^\dagger\d_i\Phi]
+i\mu q(\mu,T)\left(\Tr[\Phi^\dagger\d_0\Phi]-\Tr[\Phi\d_0\Phi^\dagger]\right)
\ . 
\ee
Since all components of the order parameter field have zero
baryonic charge, the current carries no charge. In chiral perturbation
theory, the situation is analogous since the pions do not carry baryonic charge. 
Consequently, the baryonic chemical potential does not appear in that
context.  
At the chiral phase transition, it is possible that the charge function
$q(\mu,T)$ is nonzero for $\mu\neq0$ since such a term is not
excluded by the global symmetries.  As we have shown above, however, 
the influence of $q(\mu,T)$ on the
fixed point structure is negligible as long as $T_c\gg |m|$.  

In Ref.\ \cite{HS}, Hsu and Schwetz considered the $\beta$-functions
for massless QCD at nonzero baryonic chemical potential. They
suggested that the linear derivative terms can be neglected along the
entire phase transition in the $(\mu,T)$-phase diagram.  Consequently,
they found that the order of the chiral phase transition does not
change in the $(\mu,T)$-plane.  Based on our analysis here, we agree
with their arguments as long as $T_c\gg |m|$.  For $T_c\sim m$, the
consistency of the approach breaks down and an alternative expansion
scheme must be employed. Such a theory must interpolate between the
critical behavior in the three--dimensional theory relevant for
$T_c\gg|m|$ and the full four--dimensional Landau theory relevant at
$T=0$. We are not aware of the existence of such a scheme. In addition
let us also stress that for low temperature and high baryon chemical
potential the symmetry breaking pattern is still under debate.


\section{Three--color QCD at nonzero isospin chemical potential}
\label{sec:Nc=3muI}

Finally, we discuss the influence on the fixed point structure of the
$\beta$-functions if we allow for different chemical potentials for
different flavors. A physically relevant case is given by $N_c=3$ with
two massless quarks at nonzero baryonic chemical potential,
$\mu_B=\mu_u+\mu_d$, and nonzero isospin chemical potential,
$\mu_I=\mu_u-\mu_d$. The isospin chemical potential breaks the flavor
invariance explicitly and we are therefore in a similar situation as
in Sec. \ref{sec:Nf=Nc=2}.  Chiral perturbation theory at nonzero
$\mu_I$ has been discussed in \cite{SS,SSS,KT,STV2}. Since the
$\beta$-functions and the asociated fixed-point structure have not
been studied earlier in the literature, we will describe these issues
and the results in some detail. 

At $\mu_I=0$, the flavor symmetry is $SU(2)\times SU(2)\times
U_A(1)\times U_V(1)$ and again following \cite{robfrank} we assume
that the chiral symmetry breaking order parameter field can be
parametrized by a complex $N_f\times N_f$ matrix $\Phi$ transforming as 
\be
\Phi \to U \Phi V \ ,
\ee
where $U,V\in U(N_f)$.
At nonzero isospin chemical potential, the four--dimensional Landau theory
is
\be 
L_0 & = & 
\frac{1}{2}\Tr[(\del_\nu\Phi^\dagger ) (\del_\nu\Phi)]
-\frac{\mu_I}{4}\Tr[[\tau_3,\Phi] \d_0\Phi^\dagger-h.c.]
+\frac{\mu_I^2}{4}\Tr[\Phi\tau_3\Phi^\dagger\tau_3-\Phi\Phi^\dagger] 
+ V \ ,\ee 
where the potential is given by 
\be
V = \frac{m^2}{2}\Tr[\Phi^{\dagger} \Phi]+\lambda_1(\Tr[\Phi^{\dagger}
\Phi])^2 +\lambda_2\Tr[(\Phi^{\dagger} \Phi)^2]
+c\left[\det(\Phi)+\det(\Phi^\dagger)\right] \ .
\ee
To account for the possible presence of the axial anomaly at the
chiral phase transition we have included the term with the
proportionality constant $c$. For $c\neq0$ the $U(1)$ axial invariance
is explicitly broken.  
The chiral condensate breaks $SU(2)\times SU(2)$ to $SU(2)$. The
remaining $SU(2)$ is broken explicitly when the isospin chemical
potential is nonzero.  Furthermore if $\mu_I>|m|$ then the $U_V(1)$
is spontaneously broken as a pion condensate forms preferring a 
particular direction in isospin space, say 
\be
\Phi_0\equiv i\phi_0\tau_2 \ .
\ee

As for two--color QCD at nonzero baryonic chemical potential, the isospin
chemical potential in three--color QCD splits the masses of the charged and
the uncharged modes.  At criticality, only the zeroth charged Matsubara modes
are massless. The effective three--dimensional theory at criticality 
is
\be \label{L3dmuI}
L_{\rm 3d-eff}(T_c) & = & 
\frac{1}{2}\Tr[(\del_i\Phi^\dagger ) (\del_i\Phi)]
+\lambda_1(\Tr[\Phi^\dagger \Phi])^2 
+\lambda_2\Tr[(\Phi^\dagger \Phi)^2] +c\left[\det(\Phi)+\det(\Phi^\dagger)\right] 
\ee 
where $\Phi$ now has four real components $a$, $b$, $d$, and $f$
\be
\Phi=\left(\begin{array}{cc} 0 & a+i b \\ d+if & 0 \end{array}\right) \ .
\ee
In order to determine the one loop $\beta$-functions, 
we expand the effective 3 dimensional Lagrangian about the vacuum
$\Phi\to\Phi_0+\Phi$ and make use of the background field method at
one-loop level. Ignoring the axial anomaly for the moment, the
$\beta$-functions are  
\be 
\beta_1 = &&\kappa\frac{\del\lambda_1}{\del\kappa} =
-\epsilon\lambda_1 +\frac{1}{\pi^2}\biggl [ 
6\lambda_1^2+4\lambda_1\lambda_2\biggr ]\label{betafunctions1} \ , \\
\beta_2 = &&\kappa\frac{\del\lambda_2}{\del\kappa} 
= -\epsilon\lambda_2 +\frac{1}{\pi^2}\biggl [
5\lambda_2^2+6\lambda_1\lambda_2 \biggr ] \ ,  \label{betafunctions2}
\ee
where $\kappa$ is the arbitrary mass scale.
To order $\epsilon$, the fixed points ($\lambda_1^*,\lambda_2^*$) and
the eigenvalues of the stability matrix $S$ at the fixed points are 

\begin{itemize}

\item $\lambda_1^*=\lambda_2^*=0$. \ \ \ \ \ \ \ \ \ The eigenvalues of
$S$ at $(\lambda_1^*,\lambda_2^*)$ are $-\epsilon$ and $-\epsilon$ and
hence this fixed point is not stable.    

\item $\lambda_1^*=0$, $\lambda_2^*=\epsilon\pi^2/5$. The eigenvalues of
$S$  at $(\lambda_1^*,\lambda_2^*)$ are $-\epsilon/5$ and $-\epsilon$
and hence this fixed point is not stable.   

\item $\lambda_1^*=\epsilon\pi^2/6$, $\lambda_2^*=0$. The eigenvalues of
$S$  at $(\lambda_1^*,\lambda_2^*)$ are $0$ and $\epsilon$ and hence
this fixed point is marginally stable.    

\end{itemize}

There is one marginally stable fixed point which implies that the
order of the phase transition is not determined at one-loop in the 
absence of the axial anomaly. 

The effects of the axial anomaly can be understood by power--counting
arguments.  For $N_f=2$, the anomaly term in (\ref{L3dmuI}) is a
mass-like operator.  The explicit breaking of the $U_A(1)$ then splits
the masses of the modes with the effect that only two Matsubara modes
are massless at criticality. The effective theory of these massless
modes at $T_c$ is the familiar $O(2)$ symmetric $\phi^4$ theory and 
the phase transition is of second order \cite{GZ-J}. 

Let us compare this result to the one at $\mu_I=0$ derived in
\cite{robfrank}.  For $c=0$ and $\mu_I=0$ the phase transition is
first order induced by fluctuation.  A nonzero $\mu_I$ induces a
marginally stable fixed point.  For $c\neq0$, the phase transition may
be second order. If so it is characterized by $O(4)$ critical
exponents at $\mu_I=0$ and by $O(2)$ at $\mu_I\neq0$.

Introducing the baryonic chemical potential in addition to a nonzero
isospin chemical potential does not affect the order of the chiral
phase transition.  The baryonic chemical potential does not explicitly
break any additional symmetries and in the range we consider it does
not lead to additional condensates. In conclusion: Our treatment of 3
color QCD confirms that explicit and spontaneous breaking of
symmetries induced by the chemical potential changes the fixed point
structure of the $\beta$-functions and hence the predicted order of
the phase transition can change. The combined predictions are
summarized in figure \ref{tab:all}.


\section{Conclusions}
\label{sec:conc}

We have described the effect of a chemical potential on the order of
the temperature induced phase transitions in relativistic systems. We
have focused on the stability of the fixed points of the
$\beta$-functions and have studied three examples.  The examples
studied illustrates three main effects of the chemical potential: {\sl
1)} the explicit breaking of Lorentz invariance, {\sl 2)} the explicit
breaking of global symmetries and {\sl 3)} spontaneous breaking of
symmetries through Bose condensation. The discussion is in this way
relevant for all Landau theories which are relativistically invariant
at zero temperature and chemical potential.

Our examination of the $U(1)$ model shows that the effect of Lorentz
breaking in the Landau theory for the order parameter does not affect
the renormalization group equations as long as $T\gg |m|$. The
$\beta$-functions are effected by a nonzero chemical potential only
through the breaking of internal symmetries in addition to Lorentz
invariance. We have illustrated this by examining QCD with two colors
and two flavors. The existence of fluctuation induced phase
transitions in two--color QCD is studied in further detail in
\cite{WLS}.

As another example of how the chemical potential affects the stability
of the $\beta$-functions, we considered the chiral phase transition in
ordinary three color QCD at nonzero baryonic chemical potential. In
this case the order parameter field is neutral with respect to the
charge and the Lorentz breaking in the effective theory is suppressed.
In the range of temperatures and chemical potentials under
consideration no additional symmetries are broken by the baryonic
chemical potential. In agreement with \cite{HS} we conclude that the
order of the phase transition in QCD does not change. However, we
stress that this result is only self consistent for $T_c\gg |m|$.
The situation is quite different when we consider a nonzero isospin
chemical potential in three--color QCD. In that case part of the order
parameter field does have a nonzero third component of isospin. It is
only these components which give  massless zeroth Matsubara modes in
the effective three--dimensional theory at criticality. The
modes with a zero third component of isospin remain massive at the phase
transition when the isospin chemical potential is nonzero. Since the
number of degrees of freedom at the phase transition changes the
$\beta$-functions change. This leads to a new stability pattern of the
fixed point and as such to a different prediction for the order of the
phase transition.

Finally let us stress that the analysis as performed here does not
address all caveats associated with fluctuation induced phase
transitions. For example fixed points outside the reach of
perturbation theory can change the conclusions.

\acknowledgements It is our pleasure to thank A.D. Jackson and
R.D. Pisarski for useful conversations and comments on the
manuscript. K.S.\ wishes to thank T. Sch\"afer and J.J.M. Verbaarschot
for pointing out an error at a crucial point.  
The work of K.S. was supported by the Rosenfeld foundation. 
J.T.L.\ is supported by the U.S.\ DOE grant DE-FG02-97ER41027.

\newpage

\begin{figure}
\Large
\begin{tabular}{|c||c|c|}
\hline 
 & &  \\
$\,\,\,$$N_c=2$, $\mu_B\neq0$ $\,\,\,$  & $\,\,\,$  No $U_A(1)$
 breaking $\,\,\,$ &$\,\,\,$  $U_A(1)$ breaking $\,\,\,$  \\
 & &  \\
\hline \hline 
 & &  \\
$N_f=2$ & Inconclusive & $2^{\rm nd}$-order\\
&  {\normalsize (FI $1^{\rm st}$-order at $\mu_B=0$)} & {\normalsize ($2^{\rm nd}$-order at $\mu_B=0$)} \\
\hline 
 & &  \\
$N_f = 4$ & $2^{\rm nd}$-order  & FI $1^{\rm st}$-order \\
  & {\normalsize (FI $1^{\rm st}$-order at $\mu_B=0$)} & {\normalsize (FI $1^{\rm st}$-order at $\mu_B=0$)} \\
\hline 
 & &  \\
$N_f \geq 6$  & $2^{\rm nd}$-order &  $2^{\rm nd}$-order \\
 & {\normalsize (FI $1^{\rm st}$-order at $\mu_B=0$)} & {\normalsize (FI $1^{\rm st}$-order at $\mu_B=0$)} \\
\hline
\end{tabular}
\vspace{1cm}

\begin{tabular}{|c||c|c|}
\hline
 & &  \\
$\,\,\,$ $N_c=3$, $\mu_I\neq0$ $\,\,\,\,$ & $\,\,\,$ No $U_A(1)$
 breaking$\,\,\,$  & $\,\,\,$ $U_A(1)$ breaking $\,\,\,$  \\
 & &  \\
\hline \hline 
 & &  \\
$N_f=2$ & Inconclusive & $2^{\rm nd}$-order \\
&  {\normalsize  (FI $1^{\rm st}$-order at $\mu_I=0$)} & {\normalsize ($2^{\rm nd}$-order at $\mu_I=0$)} \\
\hline
\end{tabular}
\vspace{1cm}

\normalsize 
\caption{The predicted order of the chiral phase transition in
massless QCD
with two and three colors \label{tab:all} at respectively nonzero
baryon chemical potential and nonzero isospin chemical potential.}
\end{figure}


\end{document}